\documentstyle[prb,aps,preprint,epsf]{revtex}
\def\geqap{\,\raise 2pt \hbox{$>\kern-11pt \lower 5pt \hbox{$\sim$}$}\,}
\def\leqap{\,\raise 2pt \hbox{$<\kern-10pt \lower 5pt \hbox{$\sim$}$}\,}
\begin{document}
\draft
\title{Complex orbital state in manganites}
\author{Ryo Maezono and  Naoto Nagaosa}
\address{Department of Applied Physics, University of Tokyo,
Bunkyo-ku, Tokyo 113-8656, Japan}
\date{\today}
\maketitle
\begin{abstract}
The $e_g$-orbital states with complex coefficients of the linear combination
of $x^2-y^2$ and $3z^2-r^2$ are studied for the ferromagnetic state in
doped manganites.
Especially the focus is put on the competition among uniform complex,
staggered complex, and real orbital states.
As the hole-doping $x$ increases, the real, the canted complex,
and the staggered complex orbital states appears successively.
Uniform complex state analoguous to Nagaoka ferromagnet does 
not appear.
These complex states can be expressed as a resonating state among the planer
orbitals as the orbital liquid, accompanied by no Jahn-Teller
distortion.
\end{abstract}
\narrowtext
\section{Introduction}
The role of the orbital degrees of freedom has recently attracted
considerable interests as one of the key to understand the colossal 
magneto-resistance (CMR) observed in doped manganites.
\cite{MAE98a,BRK98,END99,ISH97,OKA99,KIL99,KHA99,HOR99,MRE99a,YUN98c,SHB97,TAK98,BRT98}
The orbital state of the conduction electrons is described as a linear 
combination of two wavefunctions,
$\left| {x^2-y^2} \right\rangle$ and $\left| {3z^2-r^2} \right\rangle$,
of the degenerate $e_g$ orbitals.\cite{KUG72}
In previous studies, \cite{MAE98a,KUG72} the linear combination with only
real coefficients (real orbital state) has been considered.
This is because theories of the orbital ordering have been developed 
mainly to describe the parent compounds of CMR materials, 
\cite{KUG72,ISH96,SHI97,KOS97}
in which the static Jahn-Teller deformation is observed.
\cite{MAT70}
Such a deformation stabilizes the real orbital state and it was
reasonable to exclude the linear combination with complex 
coefficients (complex orbital state).
Recently the orbital state in {\it doped} compounds is studied
concerning the properties of CMR materials.
\cite{MAE98a,BRK98,END99,ISH97,OKA99,KIL99,KHA99,HOR99,MRE99a,YUN98c,SHB97,TAK98,BRT98}
Because the static Jahn-Teller distortion disappears in 
doped compounds, \cite{KAW95,MIT95} 
there is no reason to exclude the complex orbital state.
Actually such a complex orbital state has been recently studied.
\cite{KHO00,TAK00}
Khomskii \cite{KHO00} pointed out that the complex orbital state,
${{\left( {\left| {x^2-y^2} \right\rangle \pm i\left| {3z^2-r^2} 
\right\rangle } \right)} \mathord{\left/ {\vphantom {{\left( {\left| 
{x^2-y^2} \right\rangle \pm i\left| {3z^2-r^2} \right\rangle } \right)} 
{\sqrt 2}}} \right. \kern-\nulldelimiterspace} {\sqrt 2}}$,
provides locally isotropic hopping intensities with the same bandwidth
as the real orbital state, and might explain the isotropic properties observed
in CMR compounds.
Such a local isotropy cannot be realized with the uniform real orbital state.
\cite{MAE98a}
A staggered ordering is therefore needed to explain the observed
isotropic properties within the extent of the real orbital ordering
\cite{MAE98a}
(Another proposal is the orbital liquid state, where the local isotropy is
recovered by a quantum resonance between anisotropic orbital configurations,
$\left| {x^2-y^2} \right\rangle$, $\left| {y^2-z^2} \right\rangle$, 
and $\left| {z^2-x^2} \right\rangle$ \cite{ISH97}).
Based on the analogy to the Nagaoka ferromagnetism ($F$), Khomskii proposed
that the uniform ordering (orbital $F$) of the complex orbital state is 
more stable than the staggered one (orbital $AF$)
with real orbitals. \cite{KHO00}
Takahashi {\it et al}. investigated the possible complex orbital
ordering, motivated by the analogy to the octapole ordering
in heavy fermion systems with odd time reversal symmetry. \cite{SAK97}
They found that the {\it staggered} ordering of the complex orbital is
stable, being contrary to Khomskii's uniform one.\cite{TAK00}
\par
In this paper, we study the competitions among the uniform complex,
staggered complex, and real orbital states by using a model of CMR compounds 
taking the strong on-site repulsion and the orbital degeneracy into account.
\cite{MAE98a}
The complex orbital state is taken as,
$\cos {\left(\theta/ 2\right)}\cdot\left| 
{x^2-y^2} \right\rangle +i \sin {\left(\theta/2\right)}\cdot\left| 
{3z^2-r^2} \right\rangle$, and the whole possibility with the continuous
parameter $\theta$ is examined.
With realistic parameters, the complex orbital state 
is more stable than the real one in the moderately doped 
region ($0.25<x<0.45$).
The complex ordering changes from the canted one ($0.25<x<0.35$) into the
staggered ($0.35<x<0.45$) one due to the competition between the orbital
superexchange $AF$ and the orbital Nagaoka $F$.
The local isotropy  is also realized in this complex staggered phase,
where the band gap due to the doubled period brings 
about the energy gain exceeding the energy loss due to the narrower 
bandwidth than that of the uniform ordering with isotropic hopping.
With increasing $U/t$ toward the strong correlation limit, 
the former gain decreases whereas the latter loss increases.
The staggered ordering becomes unstable in this
limit, where the uniform orbital ordering wins.
In this case, however, the obtained uniform ordering is not the
complex one\cite{KHO00}
but the real one with $\left|{x^2-y^2} \right\rangle$.
Though the uniform complex ordering becomes more stable than
the staggered complex one, it has higher energy than that of the
real one.
In the weak correlation limit, on the other hand, Takahashi $et\ al.$
found that the normal metallic state becomes unstable toward the
the staggered complex ordering near the quarter filling,
\cite{TAK00} with increasing $U$.
When the Jahn-Teller coupling is further taken into account,
however, it is likely that the real orbital state is stabilized,
because the energy scale of the Jahn-Teller coupling becomes dominating
compared with the weak $U$, prefering the real state.
\par
The Jahn-Teller deformation which couples with the orbital degrees 
of freedom decreases in the complex canted phase and vanishes in the 
complex staggered phase, being consistent with
the observed disappearance of the deformation.
\cite{KAW95,MIT95}
This complex state can be expressed as a resonating state
among planer orbitals, as in the orbital liquid picture.
\cite{ISH97} 
When the resonance occurs with coherent correlations in time and space,
the complex orbital ordering is obtained, meanwhile that with incoherent
one corresponds to the orbital liquid state. \cite{ISH97}
These can be distinguished by experiments
detecting the spatial correlation of the orbital symmetry, such as
the anomalous X-ray scattering experiments. 
\cite{MUR98a,MUR98b}
Possibilities of the phase separation and broken time-reversal 
symmetry are also discussed.
%
\section{Results and discussions}
We employ the same model as that in the previous report \cite{MAE98a},
\begin{eqnarray}
H&=&
\sum\limits_{\sigma \gamma \gamma' \langle ij \rangle} 
{t_{ij}^{\gamma \gamma '}d_{i\sigma \gamma }^{\dagger}d_{j\sigma \gamma '}}
\nonumber \\
&-&
J_H\sum\limits_i {\vec S_{t_{2g} i}\!\cdot\! 
\vec S_{e_g i}}
\nonumber \\
&+& J_S\sum\limits_{\left\langle {ij} \right\rangle } {\vec S_{t_{2g} i}
\!\cdot\! \vec S_{t_{2g} j}} +H_{\rm on\ site} , 
\label{eq. 1}
\end{eqnarray}
where $\gamma$ [$=a(d_{x^2-y^2}), b(d_{3z^2-r^2})$] specifies the 
orbital and the other notations are standard. \cite{MAE98a}
The transfer integral $t_{ij}^{\gamma \gamma'}$ depends on the pair 
of orbitals $(\gamma, \gamma')$ and the direction of the 
bond $(i, j)$.\cite{MAE98a}
The spin operator for the $e_g$ electron is defined as 
$\vec S_{e_g i}={1 \over 2}\sum\limits_{\gamma \alpha \beta} 
 {d_{i\gamma \alpha }^{\dagger}\vec \sigma _{\alpha \beta }
d_{i\gamma \beta }}$ with the Pauli matrices $\vec \sigma$, 
while  the orbital isospin operator is defined as 
$\vec T_i={1 \over 2}\sum\limits_{\gamma \gamma' \sigma}  {d_{i\gamma \sigma }
^\dagger\vec \sigma _{\gamma \gamma '}d_{i\gamma '\sigma }}$.
\cite{MAE98a}
$J_H$ is the Hund's coupling between $e_g$ and 
$t_{2g}$ spins, and $J_S$ is 
the $AF$ coupling between nearest neighboring $t_{2g}$ spins.
$H_{\rm on\ site}$ represents the on-site
Coulomb interactions between $e_g$ electrons. Coulomb interactions
induce both the spin and orbital isospin moments, and actually
$ H_{\rm on\ site} $ can be written as 
\begin{eqnarray}
H_{\rm on\ site}
&=& -\sum\limits_i {\left( {\tilde \beta \vec T_i^2+\tilde
\alpha \vec S_{e_{g} i}^2} \right)} \ .
\label{eqn : eq2}
\end{eqnarray}
A parameter set with $t_0=t_{i,i+\hat z}^{bb}=0.72$ eV, $\tilde\alpha=8.1 t_0$,
and $\tilde\beta=6.7 t_0$ corresponds to the realistic
one being relevant to the actual manganese oxides. \cite{MAE98a}
In the path-integral quantization, we introduce the Stratonovich-Hubbard 
fields $\vec \varphi_S$ and $\vec \varphi_T$, representing the spin and 
orbital fluctuations, respectively.
With the large values of the electron-electron interactions above, 
both $\vec \varphi_S$ and $\vec \varphi_T$ are almost fully polarized.
\cite{MAE98a}
The meanfield theory corresponds to the saddle point configuration 
of $\vec \varphi_S$ and $\vec \varphi_T$.
We only consider the possibility of the complex orbital state
within a $F$-type spin alignment in the cubic cell.
\par
We assume the two sublattices, $I$ and $I\!I$, with
$F$-, $A$-, $C$- and $G$-type alignment.\cite{MAE98a}
On each site, the orbital is specified as a linear combination of the 
two degenerate orbital bases, $\left| {x^2-y^2} \right\rangle $
and $\left| {3z^2-r^2} \right\rangle $, as
\begin{equation}
\left| {\theta ,\varphi } \right\rangle =\cos {\theta  \over 2}\left| 
{x^2-y^2} \right\rangle +e^{-i\varphi }\sin {\theta  \over 2}\left| 
{3z^2-r^2} \right\rangle \ .
\end{equation}
$\left(\theta, \varphi \right)$ is the polar angle of the 
corresponding isospin $\vec T$.
In the limit of the infinite orbital polarization,
$\tilde\beta\rightarrow\infty$, the uniform orbital ordering with 
any $\left|{\theta,\varphi}\right\rangle$ takes the same bandwidth, 
$-\left({3 \mathord{\left/ 
{\vphantom {3 2}} \right. \kern-\nulldelimiterspace} 2} \right)
t_0$.
The polar angle $\left(\theta,\varphi\right)$ therefore controls only the
dimensionality of the band structure to optimize the kinetic energy gain,
leaving the bandwidth unchanged.
\par
Previous studies \cite{KHO00,TAK00} have focused on the states with
$\vec T // \hat e_y$ as the complex orbital states.
In this paper, we extend the possibility to $\vec T$ lying within
$yz$ plane ($\left| {\theta,\varphi=\pi/2 } \right\rangle$, real and pure 
imaginary coefficients) for the complex orbital state, 
whereas $\vec T$ within $zx$ plane ($\left| {\theta ,\varphi=0 } 
\right\rangle$) corresponds to the real orbital state.
This choice includes $\left| {x^2-y^2} \right\rangle$ and 
$\left| {3z^2-r^2} \right\rangle$ as the both ends.
With finite $\tilde\beta$, the generalized orbital canted structure
on two sublattices is examined.
\par
Fig. 1 shows the energy values in spin $F$ phase, optimized within the 
real and the complex orbital states, plotted as a function of 
the hole concentration $x$ (with $J_S=0$).
\begin{figure}[p]
\begin{center}
\vspace{0mm}
\hspace{0mm}
\epsfxsize=8cm
\epsfbox{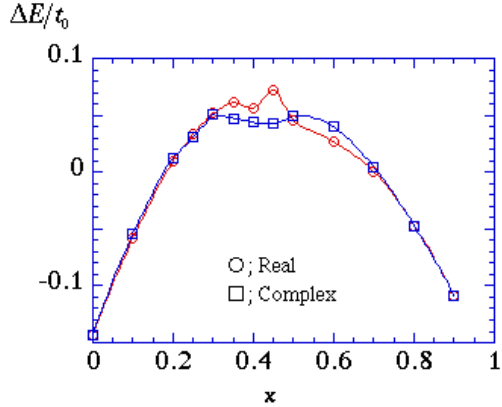}
\vspace{0mm}
\caption[aaa]{Energies of the spin $F$ phase with real and complex 
orbital states as a function of the hole concentration $x$ (with $J_S=0$).}
\label{fig : F1}
\end{center}
\end{figure}
The orbital shape specified by $\theta$ is optimized at each $x$.
The complex orbital state is realized in the moderately doped 
region ($0.25<x<0.45$).
The phase diagram as a function of $x$ and $J_S$ ($AF$
interaction between $t_{2g}$ spins) is shown in Fig. 2.
\begin{figure}[p]
\begin{center}
\vspace{0mm}
\hspace{0mm}
\epsfxsize=8cm
\epsfbox{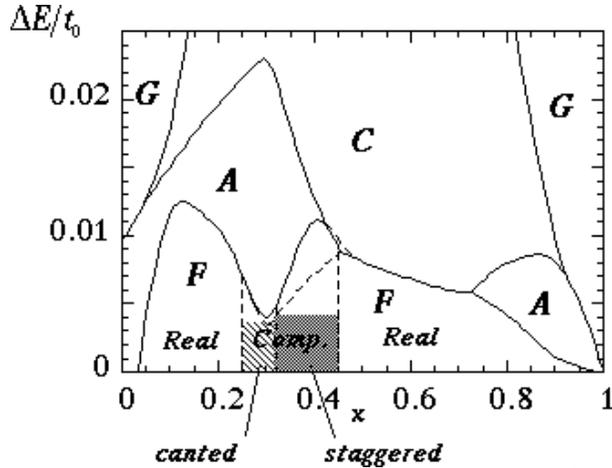}
\vspace{0mm}
\caption[aaa]{Phase diagram as a function of the hole concentration ($x$) 
and the antiferromagnetic interaction between $t_{2g}$ spins ($J_S$).
$A$, $C$, $F$, and $G$ specify the spin configuration.}
\label{fig : F2}
\end{center}
\end{figure}
In the shaded and hatched regions of the spin $F$ phase is realized the complex
orbital state.
The phase boundary depicted with a broken line is
that for the real orbital state, reported previously.\cite{MAE98a}
\begin{figure}[p]
\begin{center}
\vspace{0mm}
\hspace{0mm}
\epsfxsize=8cm
\epsfbox{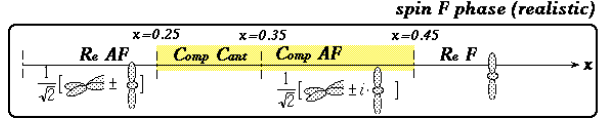}
\vspace{0mm}
\caption[aaa]{The phase diagram assuming the spin $F$ phase
with realistic parameters.}
\label{fig : F6}
\end{center}
\end{figure}
Figure \ref{fig : F6} shows the orbital phase diagram 
assuming the spin $F$ phase as a function of $x$.
The orbital ordering changes from the real staggered, the complex canted, 
the complex staggered, and to the real uniform one.
Figure \ref{fig : F3} shows the $x$-dependence of the orbital canting angle,
$\left |\theta_{I\!I}-\theta_I\right |$, for the real and the complex 
orbital states.
\begin{figure}[p]
\begin{center}
\vspace{0mm}
\hspace{0mm}
\epsfxsize=8cm
\epsfbox{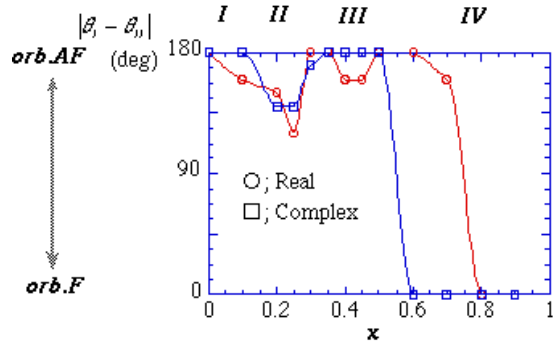}
\vspace{0mm}
\caption[aaa]{$x$-dependence of the orbital canting angle
for the real and the complex orbital states (with realistic parameters).}
\label{fig : F3}
\end{center}
\end{figure}
With increasing $x$, the canting angle once tends to take the orbital
$F$ ($I\rightarrow I\!I$), but get back to $AF$ again 
($I\!I \rightarrow I\!I\!I$).
The canted complex state with orbital $C$ is stable at $x=0.25$ and $0.35$.
With increasing $x$, the canted state changes into the staggered one for
$0.35<x<0.45$ with $\theta_I=-\theta_{I\!I}=\pi/2$ (orbital $G$)
as found in ref. 22.
We note that this $staggered$ state also gives
the locally isotropic hopping integrals,
$t_{}^x={1 \over 2}e^{-i{{2\pi } \over 3}}$,
$t_{}^y={1 \over 2}e^{i{{2\pi } \over 3}}$, and
$t_{}^z=-{1 \over 2}$.
The uniform complex state \cite{KHO00} with 
$\theta_I=\theta_{I\!I}=\pi/2$ and $t_{}^{x,y,z}=-1/2$
has higher energy.
With further doping ($I\!V$ with $x>0.5$), the orbital $F$ becomes stable 
again, but with real coefficients.
\par
\begin{figure}[p]
\begin{center}
\vspace{0mm}
\hspace{0mm}
\epsfxsize=8cm
\epsfbox{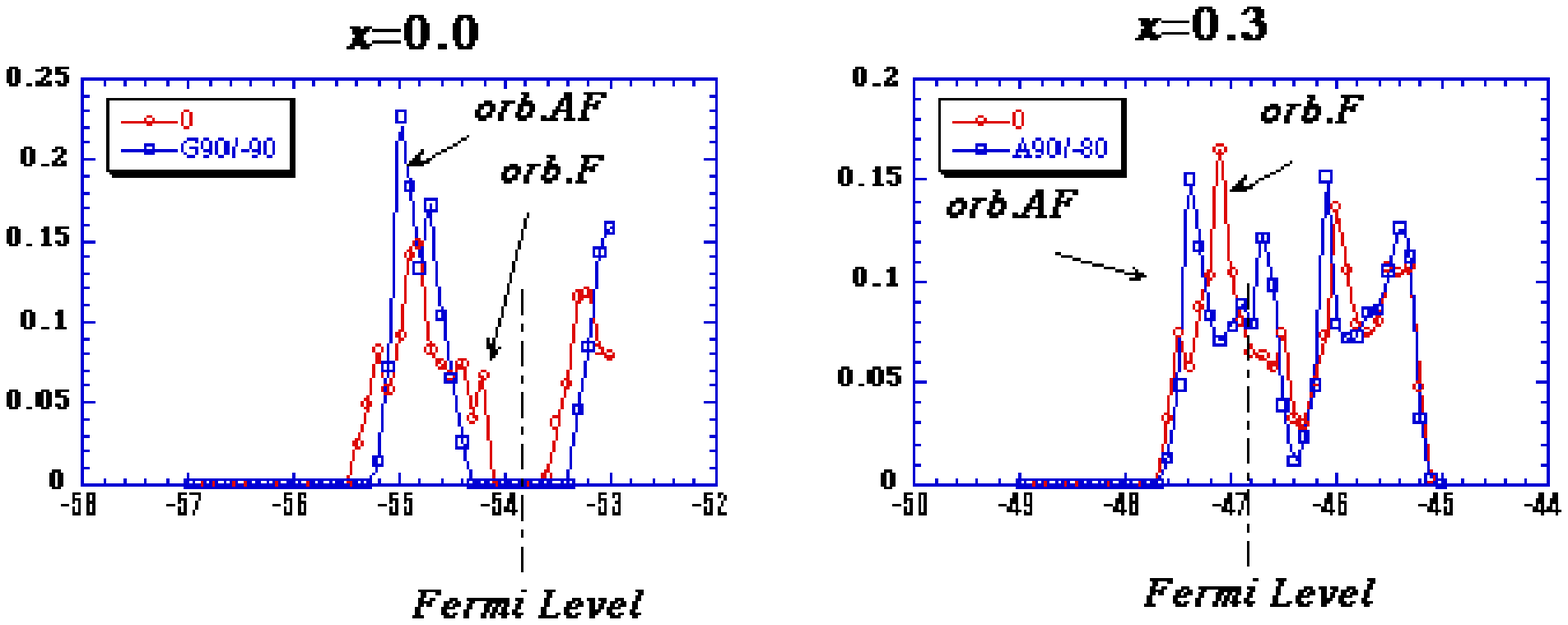}
\vspace{0mm}
\caption[aaa]{
Typical density of states (DOS) calculated with
the orbital $F$ and $AF$ ordering at $x=0$ and $0.3$.}
\label{fig : F4}
\end{center}
\end{figure}

These results can be understood as follows.
The orbital superexchange $AF$ interaction $J_{\rm AF}$ is represented by 
the shift in
the center of mass of the occupied density of states (DOS), as represented
by the Hamiltonian,
\begin{equation}
\left( {\matrix{{\varepsilon _k}&{\beta _{\rm eff}}\cr
{\beta _{\rm eff}}&{\varepsilon _{k+Q}}\cr
}} \right) \ ,
\end{equation}
with $Q=\left(\pi,\pi,\pi\right)$ being the staggered orbital
momentum.
Therefore $J_{\rm AF}$ is estimated as 
\begin{equation}
  J_{\rm AF} \cong \frac{t^2}{\beta_{\rm eff}}
\cong \frac{t^2}{\tilde\beta\left(1-x\right)} \ ,
\end{equation}
which increases as $x$ increases because $\beta_{\rm eff}$ is the constant
$\tilde\beta$ times the number of electrons $\left(1-x \right)$.
The ferromagnetic double exchange interaction $J_{\rm F}$ for the orbital
moments is represented
by the energy of the doped holes at the top of the occupied band, which
depends on the bandwidth.
The bandwidth is $t$ for the uniform ordering whereas $t^2/\beta_{\rm eff}$
for the staggered one for $t \ll \beta_{\rm eff}$ and small $x$.
$J_{\rm F}$ is therefore given as,
\begin{equation}
 J_{\rm F}\sim \left(t-\frac{t^2}{\beta_{\rm eff}}\right)\ \cdot x\ ,
\end{equation}
which represents the relative kinetic energy gain of the orbital $F$
state measuring from that of the staggered state.
It should be noted here that the notation $J_{\rm F}$ is rather symbolic, 
and the Hamiltonian is not written as $-J_{\rm F}\sum_{ij}
{\vec T_i\cdot\vec T_j}$.
Based on these considerations, the results in Fig. \ref{fig : F2}
and \ref{fig : F6} are interpreted as follows.
Here we assume the ferromagnetic spin alignment.
At $x=0$, $J_{\rm F}$ vanishes meanwhile $J_{\rm AF}$ is finite,
leading to the orbital $AF$.
With small doping, $J_{\rm F}$ becomes finite,
leading to the tendency toward the orbital $F$ seen in the 
region $I\!I$ in Fig. \ref{fig : F3}.
(This corresponds to the crossover from the orbital superexchange $AF$ to
the orbital double exchange (Nagaoka) $F$ with the doping.)
To understand the reentrant of the orbital $AF$ in the region $I\!I\!I$,
we note that ${t^2}/{\beta_{\rm eff}}={t^2}/{\tilde\beta\left(1-x\right)}$
increases as $x$, which enhances $J_{\rm AF}\left(x\right)$ and
suppresses $J_{\rm F}$.
Actually, the difference in the bandwidth of DOS between 
the uniform (orb. $F$) and the staggered (orb. $AF$) structures
is hardly seen 
at $x=0.3$ in Fig. \ref{fig : F4}, corresponding to $J_{\rm F}\sim 0$.
\noindent
The staggered ordering is therefore stabilized with increasing 
$J_{\rm AF} \left(x\right)$ in the moderately doped region.
In the heavily doped region ($I\!V$), the expression of $J_{\rm F, AF}$
does not hold any more because $t\approx\beta_{\rm eff}$.
There the staggered ordering is unstable due to the lower bandwidth
than that of the uniform one, leading to the orbital $F$ ordering.
\par
The competition between the real and the complex orbital states is
understood as follows. 
The complex state is stabilized only in the moderately doped region with
the advantage of the isotropic band structure.
In the other region, some other mechanism is rather important
than the isotropy:
In the small doping region, the real state realized in Fig. \ref{fig : F1}
and \ref{fig : F6} ($x < 0.25$) is found to be stabilized mainly 
due to the hybridization between the occupied and the unoccupied bands 
via the off-diagonal hopping integrals.
In the heavily doped region, on the other hand,
the low dimensional band structure,
$\theta_I=\theta_{I\!I}=0$ (two dimensional)  or $\pi$
(quasi-one dimensional), is prefered where the isospin moment is along the
$z$ axis (real orbital state).
This is due to the relative location between the van Hove singularity of 
DOS and the fermi level. \cite{MAE98a}
The fermi level with small electron concentration 
($x \sim 1$, heavily doped region) is located near the band edge.
The low dimensional band structure with singularities at the top and the
bottom of the band can therefore lower the kinetic energy effectively
with large accomodation at the singularity near the fermi level.
\par
\begin{figure}[p]
\begin{center}
\vspace{0mm}
\hspace{0mm}
\epsfxsize=8cm
\epsfbox{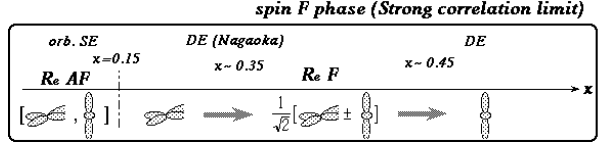}
\vspace{0mm}
\caption[aaa]{Schematic phase diagram in the strong correlation limit.}
\label{fig : F5}
\end{center}
\end{figure}

With increasing $\tilde\beta/t$, the staggered state becomes
unstable because $J_{\rm AF}$ goes to zero whereas $J_{\rm F}$
remains to be finite.
This corresponds to the recovery of the orbital Nagaoka $F$.
One can therefore expect the uniform complex state
in the small doped region with the strong correlation limit.
The obtained ordering is however the {\it real}
uniform one with $\left|{x^2-y^2}\right\rangle$, not the {\it complex} one.
The schematic phase diagram in this limit is
given in Fig. \ref{fig : F5}.
\noindent
This can be understood as follows.
In this limit, only the DOS near the band edge matters
because the bandwidth of the uniform state does not depend on the orbital 
shape.
The DOS arises at the edge most sharply with
$\left|{x^2-y^2}\right\rangle$, \cite{MAE98a}
giving the largest kinetic energy gain.
Orbital Nagaoka $F$ is therefore realized with 
$\left|{x^2-y^2}\right\rangle$ in the strong correlation limit.
(It should not be confused with the real state in the small doping region
with realistic parameters (Fig. \ref{fig : F6}), where the ordering is 
orbital $AF$ stabilized due to the inter-band hybridization.)
\par
Curves obtained in Fig. \ref{fig : F1} are non-monotonic with a common 
tangential line contacting at two different $x$, where the curve is convex 
upwards.
This means that the phase coexistence with two different concentrations has 
higher energy than the single phase.
Therefore a spontaneous phase separation \cite{MRE99a}
does not occur in our results.
\par
The $e_g$ state specified with the isospin orientation
$\vec T$ stabilizes the Jahn-Teller (JT) deformation expressed as
$\left[ {\left\langle {\vec T} \right\rangle _z\cdot Q_{u}^{}+\left
\langle {\vec T} \right\rangle _x\cdot Q_{v}^{}} \right]$, where
$Q_u$ and $Q_v$ denote the normal coordinates of the displacement
of the oxygen ions ${\Delta_\alpha}$ ($\alpha=x,y,z$): \cite{MAE98a,KAN60}
\begin{eqnarray}
Q_u^{}={{2\Delta _z-\Delta _x-\Delta _y} \over {\sqrt 6}}
\ , \ 
Q_v^{}={{\Delta _x-\Delta _y} \over {\sqrt 2}} \ .
\end{eqnarray}
The complex state realized with $0.35<x<0.45$ corresponds to 
$\vec T /\!/ \hat e_y$.
With this ordering, therefore, the JT distortion does not occur.
The observed disappearance of the JT distortion in the spin $F$ metallic region
\cite{KAW95,MIT95} might be explained by this type of the orbital ordering.
Another theoretical proposal is the orbital
liquid state where the planer orbitals, $\left| {x^2-y^2} \right\rangle$, 
$\left| {y^2-z^2} \right\rangle$, and $\left| {z^2-x^2} \right\rangle$
are resonating to form a quantum liquid state.\cite{ISH97}
With this resonance, the local isotropy is recovered, and the JT 
distortion disappears on average.
The complex state can actually be expressed in the form of such a 
resonance as,
\begin{equation}
{1 \over {\sqrt 2}}\left| {x^2-y^2} \right\rangle \pm i{1 \over {\sqrt 2}}
\left| {3z^2-r^2} \right\rangle ={{\sqrt 2} \over 3}\left[ {\left| {x^2-y^2} 
\right\rangle +e^{\pm i{{2\pi } \mathord{\left/ {\vphantom {{2\pi } 3}} 
\right. \kern-\nulldelimiterspace} 3}}\left| {z^2-x^2} \right\rangle +e^{\mp 
i{{2\pi } \mathord{\left/ {\vphantom {{2\pi } 3}} \right. \kern-
\nulldelimiterspace} 3}}\left| {y^2-z^2} \right\rangle } \right] \ .
\label{eq. 17}
\end{equation}
This can be regarded as a formation of the $T_y$-component 
by the resonance via the transverse components $T^\pm = T_x + i T_y$.
From Eq. (\ref{eq. 17}), the complex orbital ordering corresponds to
the coherent (in time and space) resonance among the planer orbitals,
with the relative phases $e^{\pm 2\pi/3}$ being fixed.
The orbital liquid state, on the other hand, corresponds to the resonance
without phase coherence.
In this sense, the complex orbital state obtained here is the meanfield
state akin to the orbital liquid  state, and may provide
a rough estimation of the energy of the latter state.
\cite{ISH97}
Because both states give no JT distortion on average,
$direct$ observations of the orbital state is needed to distinguish them,
not via the lattice deformation, but via the difference of the spatial
orbital correlations.
Several probes are available, such as the anomalous 
X-ray scattering, \cite{MUR98a,MUR98b}
the X-ray charge density study by using of the maximum entropy method (MEM),
\cite{TAK99} the magnetic Compton scattering,\cite{KOI97}
and the polarized neutron scattering \cite{AKI76}.
\par
In summary, we studied the competitions among the uniform complex,
staggered complex, and real orbital states in CMR compounds.
In the moderately doped region, the complex orbital state is
stabilized, where the ordering changes from the canted one ($x=0.25$, $0.3$) 
to the staggered one ($0.35< x<0.45$).
This can be understood in terms of the competition among 
the band narrowing and the gap associating with
the staggered structure, and the hybridization with the 
unoccupied band.
The staggered complex ordering is not accompanied with the Jahn-Teller
deformation.
The obtained complex orbital is a coherent resonance among
the planer orbitals with constant phase angles.
If the coherency is lost, the state reduces to the orbital liquid state
which can be distinguished by the observation of the spatial orbital 
correlations.
The phase separation does not occur with 
the complex orbital state obtained here.
\par
The authors would like to thank D. Khomskii, Y. Tokura
for their valuable discussions. 
This work was supported by Priority Areas Grants from the Ministry 
of Education, Science and Culture of Japan.
R.M. is supported by Research Fellowship of the Japan Society
for the Promotion of Science (JSPS) for Young Scientists.
%

%
\end{document}